\documentclass[prl, aps, floatfix, reprint, superscriptaddress]{revtex4-1}

\usepackage{url}
\usepackage{color} 
\usepackage{wrapfig}
\usepackage{epsfig}
\usepackage{enumerate}
\usepackage{amsmath}
\usepackage[colorlinks, linkcolor=blue, citecolor=blue]{hyperref}
\usepackage{setspace}
\usepackage{verbatim}
\usepackage{graphicx}
\usepackage{subfigure}
\usepackage{soul}
\usepackage{multirow}
\usepackage{bm}

\definecolor{orange}{rgb}{1.0,0.76,0.02}

\begin{document}

\title{Neural network based path collective variables for enhanced sampling\\ of phase transformations}

\date{\today}

\author{Jutta Rogal}
\email{jutta.rogal@rub.de}
\affiliation{Interdisciplinary Centre for Advanced Materials Simulation, Ruhr-Universit\"at Bochum, 44780 Bochum, Germany}
\affiliation{Department of Chemistry, New York University (NYU), New York, New York 10003, United States}
\author{Elia Schneider}
\affiliation{Department of Chemistry, New York University (NYU), New York, New York 10003, United States}
\author{Mark E. Tuckerman}
\affiliation{Department of Chemistry, New York University (NYU), New York, New York 10003, United States}
\affiliation{Courant Institute of Mathematical Sciences, New York University (NYU), New York, New York 10012, United States}
\affiliation{NYU-ECNU Center for Computational Chemistry at NYU Shanghai, 3663 Zhongshan Road North, Shanghai 200062, China}

\begin{abstract}
We propose a rigorous construction of a 1D path collective variable to sample structural phase transformations in condensed matter. The path collective variable is defined in a space spanned by global collective variables that  serve as classifiers derived from local structural units.  A reliable identification of local structural environments is achieved by employing a neural network based classification.
The 1D path collective variable is subsequently used together with enhanced sampling techniques to explore the complex migration of a phase boundary during a solid-solid phase transformation in molybdenum. 
\end{abstract}

\maketitle

Efficient sampling of high-dimensional conformational spaces represented by rough potential energy landscapes constitutes a 
significant challenge in the computational molecular sciences, particularly when different basins on the landscape
are separated by energy barriers significantly higher than $k_{\rm B}T$. 
In order to address this challenge,
various enhanced sampling techniques have been developed including accelerated molecular dynamics~\cite{Voter97,Voter98,Sorensen00,Voter2002,Perez09}, transition path sampling~\cite{Dellago98a,Dellago98b,Dellago2002,Bolhuis02}, metadynamics~\cite{Laio2002,Laio05,Laio08,Barducci08}, and (driven) adiabatic free energy dynamics (d-AFED)~\cite{Rosso02a,Rosso02b,Abrams2008} or temperature accelerated molecular dynamics (TAMD)~\cite{Maragliano06} and combinations of 
these~\cite{UFED,TASS}.
In many cases the sampling, and in essentially all cases the analysis of the resulting
high-dimensional free-energy landscapes, requires a projection onto a 
low-dimensional collective variable (CV) space. Indeed, the choice of the CVs is not always intuitive, 
but a meaningful representation in the low-dimensional space is crucial for capturing the correct mechanisms. 

Machine learning (ML) can provide a powerful approach to address the aforementioned challenges.  
The last decade has seen significant advances in the use of electronic structure calculations to train
ML potentials for atomistic simulations capable of reaching large systems sizes and long time
scales with accurate and reliable energies and forces.
More recently, ML approaches have proved useful in learning high-dimensional free-energy surfaces
(FESs)~\cite{Schneider_ML2017,Kevrekidis_2017}, and in providing a low-dimensional set
of CVs~\cite{Ming_2018,Hummer_2019}.  In such approaches, however, it is often difficult to interpret
the low-dimensional CVs that emerge from the learning procedure as they emerge
as abstract outputs of the ML model employed.
In this letter, we overcome this difficulty by exploiting an ML model
to identify local atomic structures and then using the ML output to construct 
a physically motivated one-dimensional CV.   The latter is then employed with an
enhanced configurational sampling scheme to characterise structural phase transformations in condensed matter.  
The basic idea of our approach is generally applicable to tackle different kinds of phases transformations.

A structural phase transformation can be viewed as a global change of the entire system that is 
associated with 
and driven by changes in the local structural environment around each atom (or other structural building blocks such as molecules).  
Furthermore, for each phase of interest (different crystalline structures, liquid and amorphous phases) we can define a distinct CV that quantifies 
the amount of a particular phase within the system.
These global CVs form an $n$-dimensional space of clasifiers where $n$ is the number of phases and any transformation between different structures can be described as a path in this classifier space.
For each path we can construct a one-dimensional path collective variable~\cite{Branduardi07} as a non-linear combination of the global classifier CVs.  The path CV can then be used in an enhanced sampling scheme, to project free energies, or to analyse the mechanism along the transformation. 
Our approach to derive the 1D path CV correspondingly involves three steps: first, we use a classification 
neural network (NN)~\cite{Behler11b,Geiger13} to identify the local structural environment around each atom in terms of the involved phases; next, this local information is combined into global classifier CVs, e.g. as the fraction of each structure in the system;  finally, we define a path that connects two phases in the classifier space and compute the corresponding path CV.

As an illustration of our approach, we study the solid-solid transformation between the topologically close-packed (TCP) A15 and the body-centred cubic (bcc) phase in molybdenum.   TCP phases are of particular interest in high-performance materials such a Ni-base superalloys~\cite{Reed06} as their formation significantly influences the materials properties~\cite{Rae01}.  In tungsten, the transformation between the A15 and bcc phase has recently attracted increased attention~\cite{Liu16,Barmak17} since for applications in microelectronics the formation of A15 should be avoided~\cite{Rossnagel02,Choi11} whereas in spintronic devices the A15 phase is the desired one~\cite{Pai12}.  
In a previous study the dynamics of the A15-bcc phase boundary in Mo was investigated using the adaptive kinetic Monte Carlo (AKMC) approach~\cite{Duncan16}. It was found that the phase boundary moves via collective displacements of groups of atoms through a disordered interface region which was associated with an effective barrier for the formation of a new bcc layer.  
Solid-state nudged elastic band calculations likewise indicate that the minimum energy path at $T=0$~K favors the nucleation of an interface and growth by phase boundary migration over a concerted mechanism~\cite{Xiao14}.
Here, we demonstrate how we can efficiently sample the phase space explored during phase boundary migration  by combining the 1D path CV with d-AFED/TAMD~\cite{Rosso02a,Rosso02b,Maragliano06,Abrams2008} 
and metadynamics~\cite{Laio2002,Laio05,Laio08,Barducci08} 
and characterise the free energy landscape along the A15 to bcc phase transformation.

The first step in constructing the path CV for structural phase transformations is the identification of the local structural environment.  The NN for structure classification applied in this work is based on a framework proposed by Geiger and Dellago to distinguish various polymorphs of ice~\cite{Geiger13}. 
We use a feed-forward NN where the input layer is given by a set of functions for each atom that serve as structural fingerprints.  Specifically, we use a mixture of a subset of the Behler-Parrinello symmetry functions, which were introduced to interpolate potential energy surfaces in condensed matter systems using NNs~\cite{Behler07,Behler11a}, and the Steinhardt bond order parameters~\cite{Steinhardt1983}.  Both the symmetry functions and the Steinhardt parameters are invariant with respect to rotation, translation, and the exchange of two atoms of the same element. 
To make the structure classification using the NN efficient, the number of input functions should be kept small.  In addition, the input functions should be simple and as short ranged as possible.  By combining the symmetry functions with the Steinhardt parameters we were able to reduce the number of input functions to 14, with 11 symmetry functions of two different types and three Steinhardt bond order parameters with $l=6,7,8$ (details regarding the input functions are given in the Supplemental Material~\cite{supplemental}).   For comparison, 45-50 symmetry functions of four different types were used in Ref.~\onlinecite{Geiger13}.  
Use of a smaller subset of the full set of input functions, {\it i.e.,} only the symmetry functions or only the
Steinhardt parameters, does not provide enough information as input to NN, 
thereby degrading the accuracy of the structure classification.

\begin{figure}
\begin{center}
\includegraphics[width=0.48\textwidth]{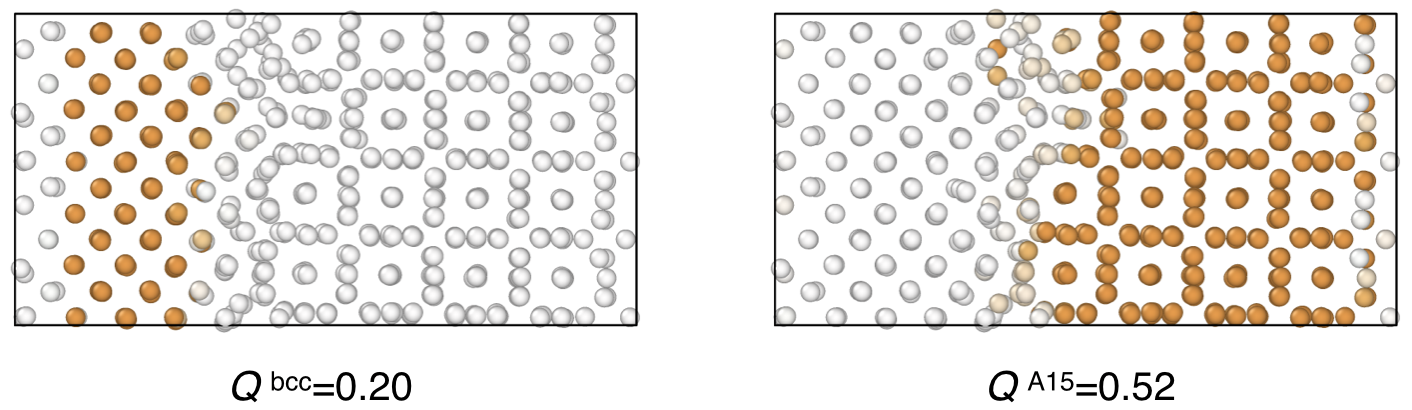} 
\end{center}
\caption{\label{fig:nnclass}
Initial interface setup after equilibration at $T=300$~K; atoms are coloured according to their local structural order parameter value determined using the NN for $q_i^{\text{bcc}}$ (left) and $q_i^{\text{A15}}$ (right) with a colour scheme from white ($q_i^j = 0$) to orange ($q_i^j = 1$).  $Q^{\text{bcc}}$ and $Q^{\text{A15}}$ are the values of the global CV defined in Eq.~\eqref{eq:globalCV}.
}
\end{figure}
The output of the NN is a vector $\mathbf{q}_i$ for each atom $i$ with one component $q_i^j$ for each of the structures $j$ of interest with $q_i^j \in [0,1]$.  In the present study, this includes bcc and A15 as well as face-centred cubic (fcc), hexagonal close packed (hcp), and a disordered structure (dis).  The disordered phase comprises local structural environments that significantly deviate from a well-ordered crystal phase and would be characteristic of amorphous or liquid phases.  In the A15-bcc phase boundary migration, this also pertains to the disordered interface region between the two crystal bulk phases.
We have used both normalised and unnormalised vectors $\mathbf{q}_i$, but in the present application, we did not observe 
any noticeable differences.
The NN was trained using snapshots from molecular dynamics (MD) simulations at different temperatures and for the different phases.  Additional snapshots were taken from dynamical simulations of a supercell containing atoms in the A15, bcc, and interface region.  In total 346,436 local atomic environments were used to train the weights of the NN (further details are given in the Supplemental Material~\cite{supplemental}).
In Fig.~\ref{fig:nnclass} the initial setup of the bcc-A15 interface is shown where the atoms are coloured according to the output of the NN for $q_i^{\text{bcc}}$ (left) and $q_i^{\text{A15}}$ (right).  The NN clearly identifies the two crystalline regions.

The second step in constructing the path CV consists in using 
the atomic output vectors of the NN to define global classifier CVs for each phase.  
In particular, for phase boundary migration, we define a global CV as the average over the local phase classification. 
For example, for bcc the global CV is
\begin{equation}
\label{eq:globalCV}
Q^{\text{bcc}} = \frac{1}{N} \sum_{i=1}^N q_i^{\text{bcc}} \quad ,
\end{equation}
where $N$ is the number of atoms in the simulation cell. Classifier CVs for the other phases, $Q^{\text{A15}}$, $Q^{\text{fcc}}$, $Q^{\text{hcp}}$, and $Q^{\text{dis}}$ are similarly obtained.
Within this definition the classifier CVs describe the phase fractions for any given configuration.

\begin{figure}
\begin{center}
\includegraphics[width=0.48\textwidth]{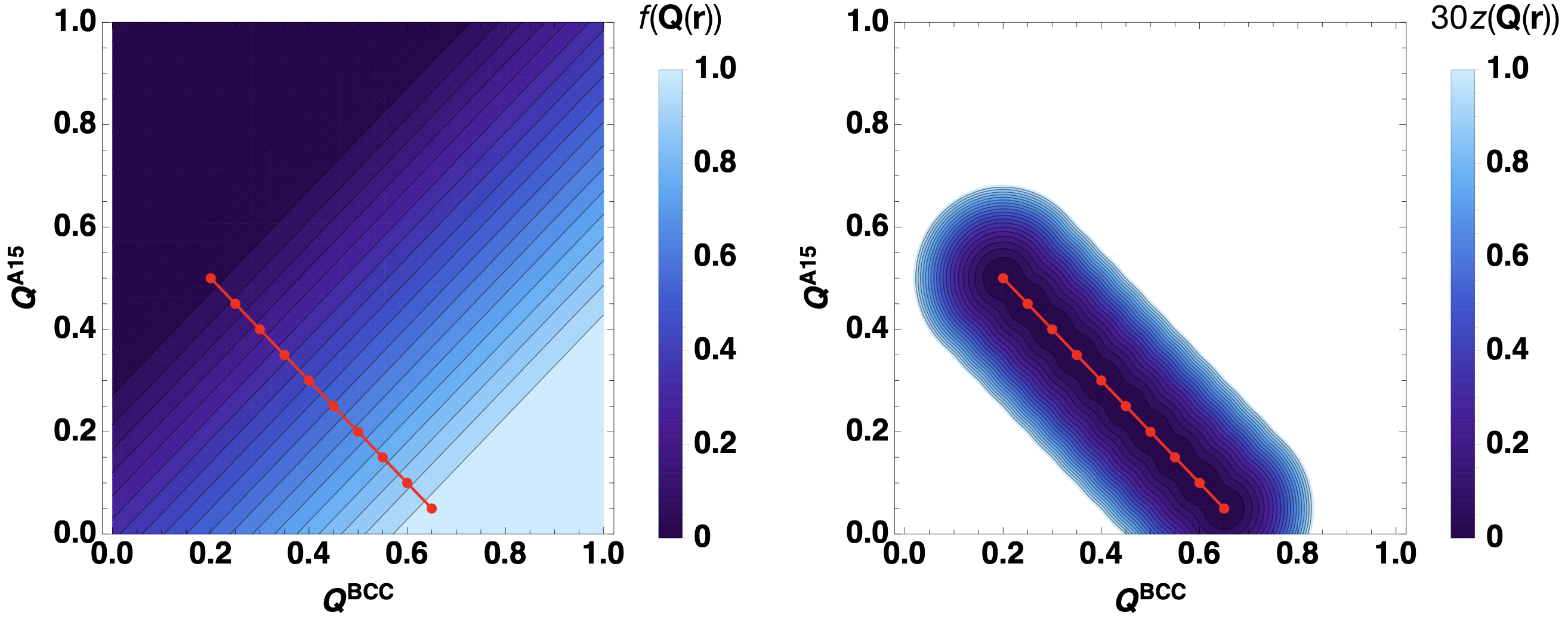}
\end{center}
\caption{\label{fig:pathcv}
Left: path collective variable $f(\mathbf{Q}(\mathbf{r}))$ in the $Q^\text{bcc}-Q^\text{A15}$ space. Right: distance function $z(\mathbf{Q}(\mathbf{r}))$ (multiplied by a factor of 30)  in the $Q^\text{bcc}-Q^\text{A15}$ space.  The red lines indicate the path along the nodal points.
}
\end{figure}
The final step consists in constructing a path CV in the space spanned by the global classifier CVs.  
In the case of the phase transformation between bcc and A15, we choose a path in the $Q^{\text{bcc}} - Q^{\text{A15}}$ space where the sum of the phase fractions is constant.  The path CV is defined as~\cite{Branduardi07}
\begin{equation}
\label{eq:pathcv}
f(\mathbf{Q}(\mathbf{r})) = \frac{1}{P-1} \frac{\sum_{k=1}^P (k-1) \exp(-\lambda|\mathbf{Q}(\mathbf{r}) - {\bm Q}_k|^2|}{\sum_{k=1}^P \exp(-\lambda|\mathbf{Q}(\mathbf{r}) - {\bm Q}_k|^2)} \quad ,
\end{equation}
where $\mathbf{Q}(\mathbf{r}) 
= \{Q^{\text{bcc}}, Q^{\text{A15}} \}$ is the position in the $Q^{\text{bcc}} - Q^{\text{A15}}$ space, ${\bm Q}_k$, $k = 1,...,P$ are the $P$ nodal points along the path, and $|\mathbf{Q}(\mathbf{r}) - {\bm Q}_k|^2$ is the square distance from the path in classifier space.  
The path has $P=10$ equidistant points and follows the diagonal from ${\bm Q}_1 = \{0.2, 0.5\}$  to $\bm{Q}_{10} = \{0.65, 0.05\}$.  In order to provide a meaningful representation of the path CV in its discretised form in Eq.~\eqref{eq:pathcv}, $\lambda$ should approximately be set to the inverse of the square distance between consecutive nodal points, yielding $\lambda=200$.  The value of the path CV together with the path in the $Q^{\text{bcc}} - Q^{\text{A15}}$ space is shown in the left graph of Fig.~\ref{fig:pathcv}.  It smoothly increases from 0 to 1 as the path fraction of bcc increases and the one of A15 decreases.  Perpendicular to the path the value of the path CV is constant.  
In addition, the function~\cite{Branduardi07}
\begin{equation}
\label{eq:distancecv}
z(\mathbf{Q}(\mathbf{r})) = - \frac{1}{\lambda} \ln\left(\sum_{k=1}^P \exp(-\lambda(\mathbf{Q}(\mathbf{r}) - {\bm Q}_k)^2)  \right)
\end{equation}
can be used
as a distance measure from the path.  $z(\mathbf{Q}(\mathbf{r}))$ can either be used as additional 
biasing coordinate or as a restraining potential~\cite{Margul_2018}.  
In the right graph of Fig.~\ref{fig:pathcv} $z(\mathbf{Q}(\mathbf{r}))$ (multiplied by a factor of 30)
is shown for $\lambda=1000$~\cite{commentlambda}.

\begin{figure}
\begin{center}
\includegraphics[width=0.45\textwidth]{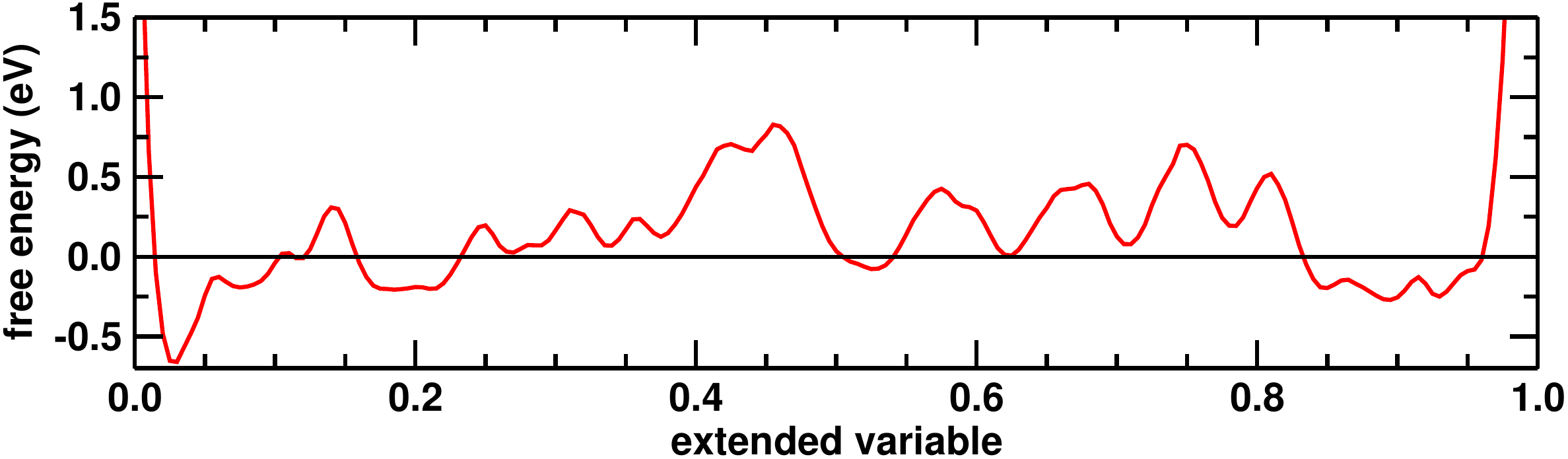}  \\[2ex]
\includegraphics[width=0.45\textwidth]{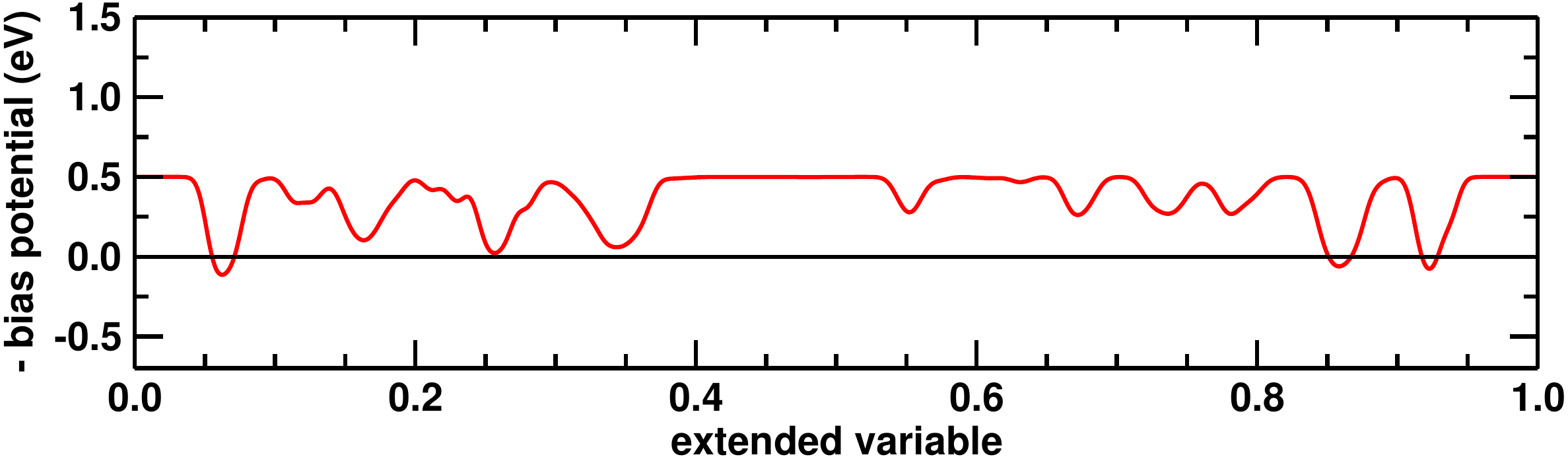} 
\end{center}
\caption{\label{fig:dafedpmf}
Top: free energy profile along the extended variable~$s$ at $T=300$~K extracted from the potential of mean force of the d-AFED simulations with $T_s = 3000$~K.
Bottom: negative of the bias potential from the metadynamics simulations.
}
\end{figure}

The path CV defined in Eq.~\eqref{eq:pathcv} is subsequently employed in d-AFED/TAMD simulations  to enhance the sampling of the bcc-A15 phase transformation.  In d-AFED~\cite{Abrams2008}, specifically, the physical system is coupled to an extended variable $s$ through a harmonic potential where the Hamiltonian in the extended phase space with a single variable $s$ 
and conjugate momentum $p_s$ is given by
\begin{equation}
\label{eq:Hdafed}
\tilde{H}(\mathbf{r},\mathbf{p},s,p_s) = H(\mathbf{r},\mathbf{p}) + \frac{p_s^2}{2 \mu_s} 
+ \frac{1}{2} \kappa (f(\mathbf{Q}(\mathbf{r})) - s)^2 \quad .
\end{equation}
Here, $H(\mathbf{r},\mathbf{p})$ is the Hamiltonian of the unperturbed system, $\mu_s$ is the ``mass'' 
of the extended variable, $\kappa$ is the harmonic coupling constant, 
and $f(\mathbf{Q}(\mathbf{r}))$ is the corresponding path CV in Eq.~\eqref{eq:pathcv}.  
In order to ensure that the free energy profile $F(s)$ along the extended variable
is correctly generated, an adiabatic decoupling between the physical and extended variables is 
needed~\cite{Abrams2008} and is achieved by choosing a large value of the mass parameter $\mu_s$. 
In this adiabatic limit, 
the sampling is accelerated by running the dynamics of the extended variable $s$ at a high temperature $T_s \gg T$.
\begin{figure}
\begin{center}
\includegraphics[width=0.45\textwidth]{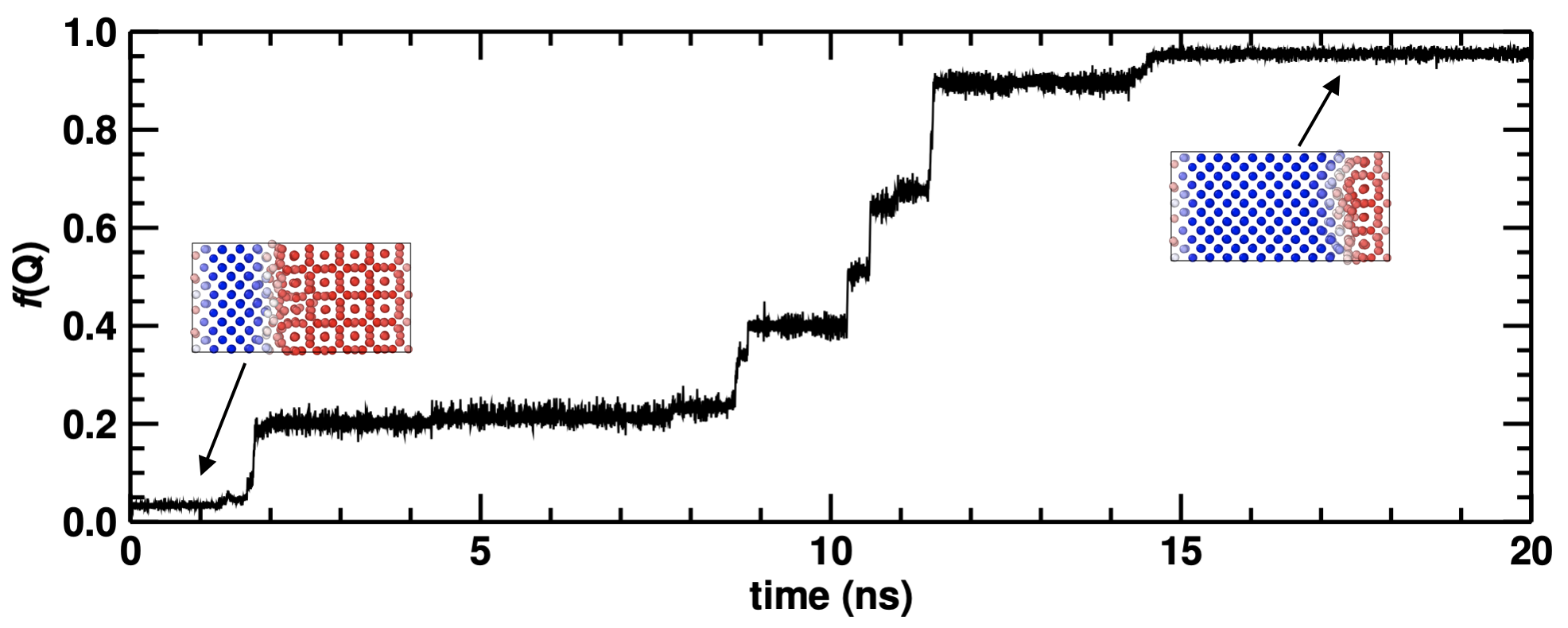} 
\end{center}
\caption{\label{fig:trajdafed}
Path collective variable $f(\mathbf{Q}(\mathbf{r}))$ as a function of time during a d-AFED run.  As the bcc phase grows the path CV increases from 0 to 1. Blue atoms are in a  bcc, red atoms  in an A15 environment.
}
\end{figure}
In Fig.~\ref{fig:trajdafed} the evolution of the path CV is shown for a representative d-AFED run in the NVT ensemble with a system temperature of $T=300$~K, an extended variable temperature of $T_s = 3000$~K, $\mu_s/\bar{\mu}_{\text{eff}} \approx 1000$ (where $\bar{\mu}_{\text{eff}}$ is the average effective mass of the path CV~\cite{Cuendet14}),  and $\kappa = 2 \times 10^5$~eV (further details are given in the Supplemental Material~\cite{supplemental}).
The entire A15 phase is transformed into bcc in less than 20~ns whereas in the unbiased system the transformation usually takes tens of microseconds~\cite{Duncan16}, emphasising the significant speed-up in the exploration of the phase space achieved by d-AFED. 
From the simulations we can estimate the free energy along the extended variable, $F(s)$, from the mean force on $s$ due to the coupling to the physical system
\begin{equation}
\label{eq:pmf}
\frac{\partial F(s)}{\partial s} = - \langle \kappa (f(\mathbf{Q}(\mathbf{r})) - s) \rangle_s^{\text{adb}}   \quad .
\end{equation}
The free energy profile for the A15 to bcc transformation extracted from 50  d-AFED runs is shown in the top graph of Fig.~\ref{fig:dafedpmf}.
The profile does not correspond the equilibrium free energy surface but rather reflects the one-way transition from A15 to bcc.  An observation of the reverse transition is not possible within this simulation cell due to the large interface mismatch between bcc and A15 (-6.73\%) and corresponding large energy difference as well the overall large pressure in the small simulation cell ($\approx 200$~kbar).  This particular system setup was chosen as a test case to compare our approach to previous results, where an effective barrier of $\Delta E_{\text{layer}} = 0.47 \pm 0.07$~eV for the formation of a bcc layer was extracted from AKMC simulations~\cite{Duncan16}.  The free energy profile in Fig.~\ref{fig:dafedpmf} clearly reflects the layer-by-layer transition observed in this system, and the respective energy barriers are $\approx 0.5$~eV, which is comparable to the effective barrier in Ref.~\onlinecite{Duncan16}.  Deviations for small and large values of $s$ are expected since at small values the interface needs to equilibrate from its initial configuration, and at large values, interactions with the second, fixed interface in the simulation cell become more pronounced. Overall the agreement with previous results is very good.


In order to test the robustness of our 1D path CV, we have also used it in metadynamics simulations.  In metadynamics~\cite{Laio2002,Laio08} a time-dependent bias potential as a function of the CV is added to the Hamiltonian, usually  in terms of Gaussians, that accelerates the exploration of the phase space by gradually filling up the wells of the energy minima.  When all minima are filled, the corresponding free energy surface becomes \emph{flat} and the system exhibits a diffusive behaviour along the CV.  The inverse of the bias potential can be used as an estimator for the free energy.  In our metadynamics simulations, we again do not converge to the equilibrium free energy surface but only obtain a rough first estimate for the one-way transition from A15 to bcc.  A representative free energy profile $F(s) = - V^{\text{bias}}(s)$ is shown in the bottom graph of Fig.~\ref{fig:dafedpmf} (details concerning the metadynamics simulations are given in the Supplemental Material~\cite{supplemental}).  
The shape again clearly indicates a layer-by-layer transition and the corresponding energy barriers are close to the expected value of 0.5~eV.  Since the formation times for a new bcc layer follow an exponential distribution typical of a rare event, the gradual filling of energy minima in the metadynamics simulation in only one direction will correspondingly result in a range of barriers including even some flat parts along the path CV.  In the d-AFED simulations we can average the mean force in Eq.~\eqref{eq:pmf} over multiple runs where each run contributes to an extensive sampling of all degrees of freedom perpendicular to $s$.

In a less constrained system, where the reverse phase transformation is also possible, we can explore both the forward and backward transition along the 1D path CV.  As an example, we have chosen the same system as above, but  the simulations are now performed at constant (zero) pressure.  In addition to the bias potential from the metadynamics simulations, we apply a restraining potential on the distance from the path CV $z(\mathbf{Q})$ in Eq.~\eqref{eq:distancecv} with $V^{\text{rest}}(z)  =  k ((z-z_0)/\epsilon)^2$, where $z_0 = 0.0$, $k=30$~eV, and $\epsilon = 0.05$. 
\begin{figure}
\begin{center}
\includegraphics[width=0.45\textwidth]{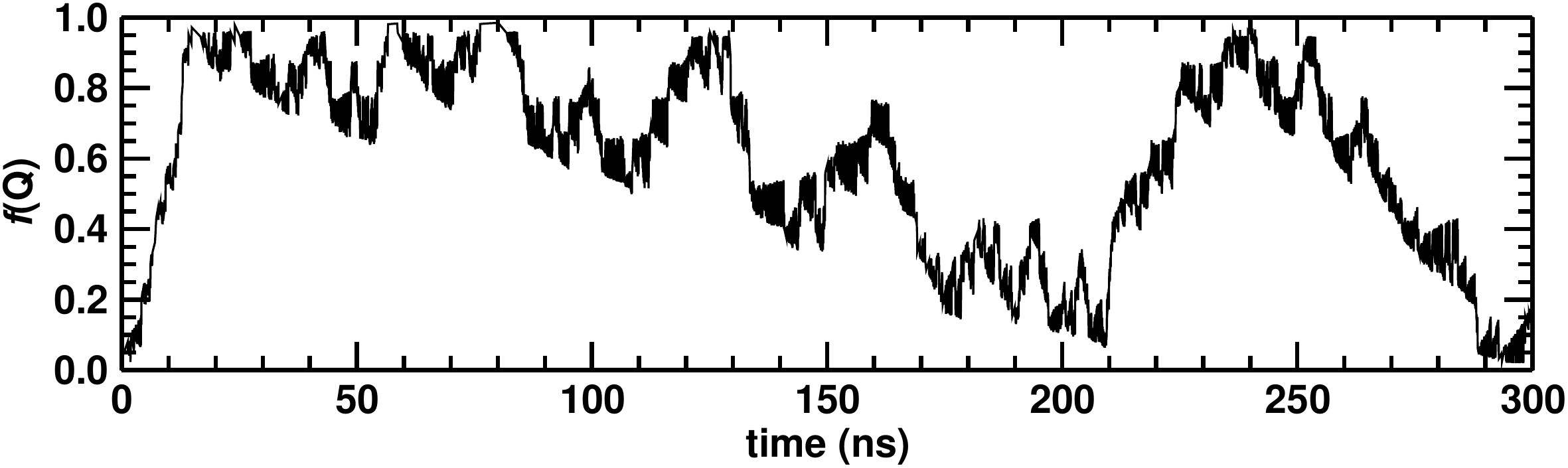}  \\[2ex]
\includegraphics[width=0.45\textwidth]{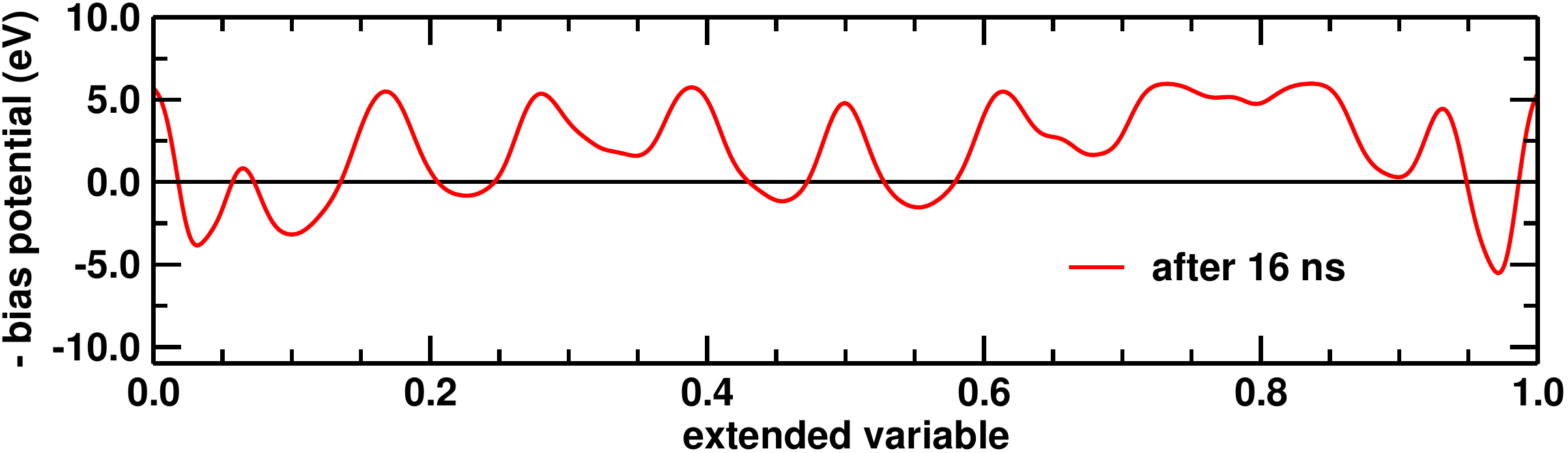}  \\[2ex]
\includegraphics[width=0.45\textwidth]{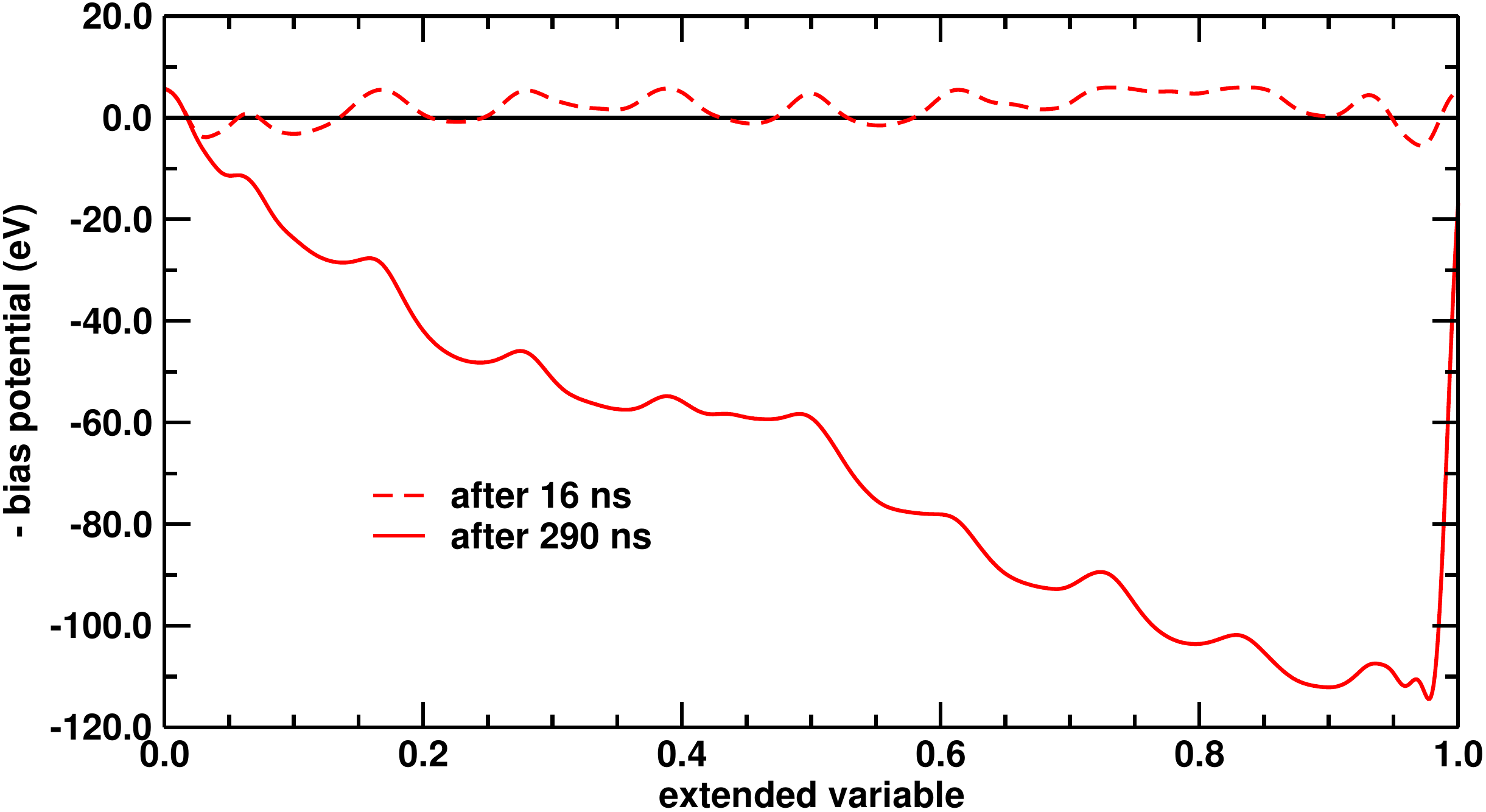} 
\end{center}
\caption{\label{fig:metanpt}
Top:  evolution of the path CV $f(\mathbf{Q}(\mathbf{r}))$ during the NPT metadynamics simulation at $T=300$~K. 
Middle: negative of the bias potential after 16~ns. 
Bottom: negative of the bias potential after 16 and 290~ns.
}
\end{figure}
The evolution of the path CV during the simulation is shown in the top graph of Fig.~\ref{fig:metanpt}.  
Up to about 16~ns the path CV increases monotonically to its maximum value indicating the first, complete A15 to bcc transformation.  The corresponding energy profile (negative of the bias potential) is depicted in the middle graph of Fig.~\ref{fig:metanpt}, again exhibiting the characteristic shape of a layer-by-layer transformation.  The energy barriers are much larger ($\approx 5$~eV) than in the previous case, which can be attributed to a decreased driving force 
(as qualified by the potential energy difference between the two phases)
and interface mobility at constant $P=0$, as well as to the additional restraining potential.  
As the bias potential continues to build up, the system starts to show a diffusive behaviour along the path CV.  Between $210-290$~ns the system transforms smoothly from A15 to bcc and back to A15. Observing this behaviour was only possible when employing the unique 1D path CV suggested here.  
Although we did not carry out constant-pressure d-AFED simulations,
we expect the same behavior would be obtained~\cite{Crystal_dAFED}.
In developing this new approach, we initially investigated several other 1D CVs,
and we also tried to sample the 2D CV space of phase fractions directly, 
but none of these simulations achieved a full sampling of the forward and backward phase transformations.     
The free energy profile after 290~ns is shown in the bottom graph of Fig.~\ref{fig:metanpt}.  
Clearly, the bcc phase is much lower in energy within this particular supercell setup and the barrier for the transformation from bcc to A15 is a factor of four larger than for the A15 to bcc transition.  This is consistent with our unbiased MD simulations, where even for very high temperatures, the reverse transition was never observed.  
For a quantitative analysis of the interface migration barriers, simulations cells with a smaller mismatch and a larger interface area will be considered in future studies.


In summary, we have proposed a general approach how to rigorously construct a 1D path CV for sampling phase transformations.   The method is based on a local structure classification using a neural network and this local information is then combined into global classifier CVs for each phase of interest.  Finally, the phase transformation can be described as a path in the $n$-dimensional classifier space and the corresponding 1D path CV can be computed as a non-linear combination of the global 
classifier  CVs.
We have shown the applicability of our approach to the solid-solid phase transformation between the bcc and A15 phases, 
which proceeds via phase boundary migration characterised by complex atomic rearrangements at the interface. Combining the 1D path CV with enhanced sampling techniques, we were able to estimate the energy profile along the transformation and, for the first time, simulate the growth of the A15 phase from the bcc phase.
Due to the computational efficiency in exploring the phase space, our approach enables the study of phase transformations in much larger simulation cells where it will be possible to consider realistic interface mismatches and transformation mechanisms including the formation of a step and growth along the step edges.


\begin{acknowledgments}
J.R. gratefully acknowledges financial support by the Alexander von Humboldt Foundation.
M.E.T. acknowledges support from the National Science Foundation partially through the
Materials Research Science and Engineering Center (MRSEC) program
DMR-1420073 and partially through CHE-1565980.
\end{acknowledgments}



\begin{thebibliography}{43}
\bibitem[{com()}]{commentlambda}
\bibinfo{note}{For the $z(\mathbf{Q})$ function a slightly larger $\lambda$
  value was chosen to avoid negative values of $z(\mathbf{Q})$ close to the
  nodal points; the overall shape of $z(\mathbf{Q})$ remains unaffected by
  increasing $\lambda$ from 200 to 1000.}
\end{thebibliography}
\end{document}


\title{Neural network based path collective variables for enhanced sampling of phase transformations -- Supplemental Material
}

\date{\today}

\author{Jutta Rogal}
\email{jutta.rogal@rub.de}
\affiliation{Interdisciplinary Centre for Advanced Materials Simulation, Ruhr-Universit\"at Bochum, 44780 Bochum, Germany}
\affiliation{Department of Chemistry, New York University (NYU), New York, New York 10003, United States}
\author{Elia Schneider}
\affiliation{Department of Chemistry, New York University (NYU), New York, New York 10003, United States}
\author{Mark E. Tuckerman}
\affiliation{Department of Chemistry, New York University (NYU), New York, New York 10003, United States}
\affiliation{Courant Institute of Mathematical Sciences, New York University (NYU), New York, New York 10012, United States}
\affiliation{NYU-ECNU Center for Computational Chemistry at NYU Shanghai, 3663 Zhongshan Road North, Shanghai 200062, China}

\maketitle

\section{Neural Network for Structure Classification}
For the structure identification we use a feed forward neural network (NN) similar to the approach proposed in Ref.~\onlinecite{Geiger13}.  For each atom a number of functions is computed that represent the local structural environment, and the output of the NN classifies the local atomic structure in terms of different crystalline or amorphous phases. The NN has 14 input nodes, 2 hidden layers with 25 nodes each, and an output layer with 5 nodes.  The 14 input nodes consist of 11 symmetry functions~\cite{Behler11a} and 3 Steinhardt local bond-order parameters.~\cite{Steinhardt1983} 
The Steinhardt parameters $q_l (i)$ for each atom $i$ are calculated according to
\begin{equation}
q_l (i) = \sqrt{\frac{4\pi}{2l + 1} \sum_{m=-l}^l |q_{lm}(i)|^2}
\end{equation}
and 
\begin{equation}
q_{lm}(i) = \frac{\sum_i^N Y_{lm}(\mathbf{r}_{ij}) f_c(\mathbf{r}_{ij})}{\sum_i^N f_c(\mathbf{r}_{ij})} \quad ,
\end{equation}
where $N$ is the total number of atoms, $Y_{lm}(\mathbf{r}_{ij})$ are the spherical harmonics, and $f_c(\mathbf{r}_{ij})$ is a cutoff function
\begin{equation}
\label{eq:fc}
f_c(\mathbf{r}_{ij}) = 
\begin{cases}
1    &    \text{if  }     |\mathbf{r}_{ij}| \leq r_\text{min} \\
\frac{1}{2} \left(\cos\left[\frac{(|\mathbf{r}_{ij}|-r_\text{min}) }{(r_\text{max}-r_\text{min})} \pi \right]+1\right)   &  \text{if  }  r_\text{min} <  |\mathbf{r}_{ij}| \leq r_\text{max} \\
0    &    \text{if  }  |\mathbf{r}_{ij}| > r_\text{max}
\end{cases}
\end{equation}
As input to the NN $q_l$ values with $l=6,7,8$  are used.  The values for the cutoff function are set to  $r_\text{min} = 3.8$~\AA~and $r_\text{max} = 4.0$~\AA.

Two different types of symmetry functions are used 
\begin{equation}
G_2 (i) = \sum_{j \neq i} e^{-\eta (|\mathbf{r}_{ij}| - R_s)^2} f_c(\mathbf{r}_{ij})
\end{equation}
and
\begin{equation}
G_3 (i) = \sum_{j \neq i} \cos(\kappa |\mathbf{r}_{ij}|) f_c(\mathbf{r}_{ij}) \quad .
\end{equation}
The corresponding values of $R_s$, $\eta$, and $\kappa$ are given in Tab.~\ref{tab:sfg}.
For all symmetry functions the cutoff function in Eq.~\eqref{eq:fc} was used with $r_\text{min} = 6.2$~\AA~and $r_\text{max} = 6.4$~\AA.
To determine the parameters of the symmetry functions snapshots were taken from molecular dynamics (MD) simulations of face-centred cubic (fcc), body-centred cubic (bcc), hexagonal close packed (hcp), the topologically closed packed A15 phase, and liquid Mo to compute histograms of the corresponding values of $G_2$ and $G_3$  for a series of $R_s$, $\eta$, and $\kappa$ values.  Subsequently, we determined the overlap of the histograms for the different structures and chose parameters for $G_2$ and $G_3$ that  resulted in  small overlaps of the distributions.

The 5 output nodes of the NN correspond to a classification value $q^j \in [0,1]$ for the five considered structures $j=$bcc, fcc, hcp, A15, and dis, where \emph{dis} characterises liquid and amorphous environments. 

\begin{table*}
\caption{\label{tab:sfg}
Parameters for the symmetry functions.  For all symmetry functions the cutoff function in Eq.~\eqref{eq:fc} was used with $r_\text{min} = 6.2$~\AA~and $r_\text{max} = 6.4$~\AA.
}
\begin{ruledtabular}
\begin{tabular}{lccp{0.5cm}lc}
$G_2$    &    $R_s$ (\AA)      &    $\eta$ (\AA$^{-2}$)     &  &  $G_3$  &  $\kappa$ (\AA$^{-1}$)  \\[0.5ex] \hline  
1                  &     2.8           &    20.0          &  &  9    &  3.5     \\
2                 &     3.2           &    20.0           &  &  10  &  4.5        \\
3                 &     4.4           &    20.0           &  &  11  &  7.0       \\
4                  &     4.8           &    20.0            \\
5                  &     5.0           &    20.0             \\
6                  &     5.3           &    20.0              \\
7                  &     5.7           &    20.0               \\
8                  &     6.0           &    20.0               \\
\end{tabular}
\end{ruledtabular}
\end{table*}
%


The initial fit of the  weights of the NN was performed using a total number of $176\,000$ atomic environments extracted from bulk and interface configurations.  The configurations were taken from MD simulations of bcc, fcc, hcp, and A15 bulk systems at $T$=300, 600, 450, and 1000~K as well as liquid at $T$=3000, 4000, and 5000~K.  The interface configurations were taken from initial driven adiabatic free energy (d-AFED)~\cite{Abrams2008} simulations (see Sec.~\ref{sec:simdafed}).  The initial set included $36\,000$ atomic environments from bcc bulk configurations, $36\,000$ from A15, $12\,000$ from fcc, $12\,000$ from hcp, $30\,000$ from liquid (classified as \emph{dis}), and $50\,000$ from interface configurations comprising bcc, A15, and \emph{dis} atomic environments.  
The training set was subsequently extended by $170\,426$ atomic environments extracted from interface configurations of further test simulations including $37\,410$ A15, $101\,371$ dis, and $31\,645$ fcc/hcp environments, that is the final training set consisted of $346\,426$ atomic environments. 
The test set consisted of $125\,000$ atomic environments ($25\,000$ for each bulk structure and the liquid) and the correct structure was assigned by the NN for more than 99\% of the test points.  For atomic environments extracted from interface configurations containing A15, BCC, and dis phases, the accuracy was more than 93\%.  The NN is thus capable of assigning the correct atomic structure with high accuracy, even in composite systems including two different crystal phases as well as a disordered interface region.

\section{Simulation setup}
\subsection{Molybdenum A15-bcc interface}
\begin{figure}
\begin{center}
\includegraphics[width=0.38\textwidth]{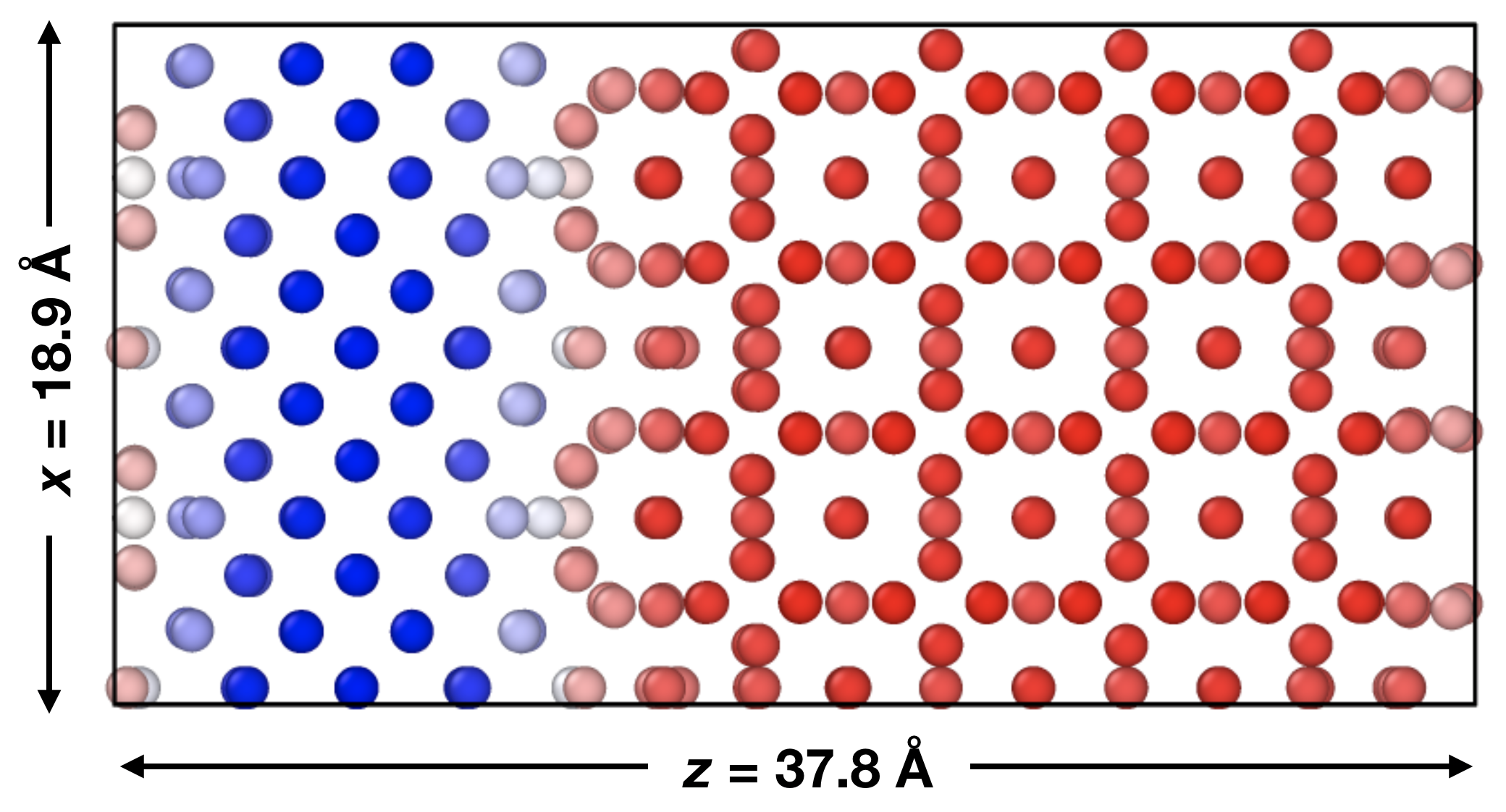} 
\end{center}
\caption{\label{fig:interface}
Simulation cell of the initial bcc-A15 interface setup, bcc coordinated atoms are coloured in blue, A15 coordinated atoms are shown in red.
}
\end{figure}

All simulations were performed using an embedded atom method (EAM) potential for molybdenum~\cite{zho01}.  The initial supercell setup is shown in Fig.~\ref{fig:interface} and is equivalent to the setup in Ref.~\onlinecite{Duncan16}.
The interface corresponds to a [100] bcc $||$ [100] A15 phase boundary with $6 \times6$ bcc and $4 \times 4$ A15 unit cells in the $xy$-plane.  In the constant volume simulations the supercell dimensions were fixed at $x=y=18.9$~\AA{} and $z=37.8$~\AA{} which corresponds to the equilibrium lattice dimensions of the bcc phase in the $xy$-plane and a compression of 7\% of the A15 phase.  This results in a potential energy difference of 0.36~eV/atom and a corresponding driving force for the formation of bcc.  At $P=0$ the potential energy difference between the two phases is only 65~meV/atom. 
%

\subsection{Driven adiabatic free energy dynamics}
\label{sec:simdafed}
\begin{figure}
\begin{center}
\includegraphics[width=0.44\textwidth]{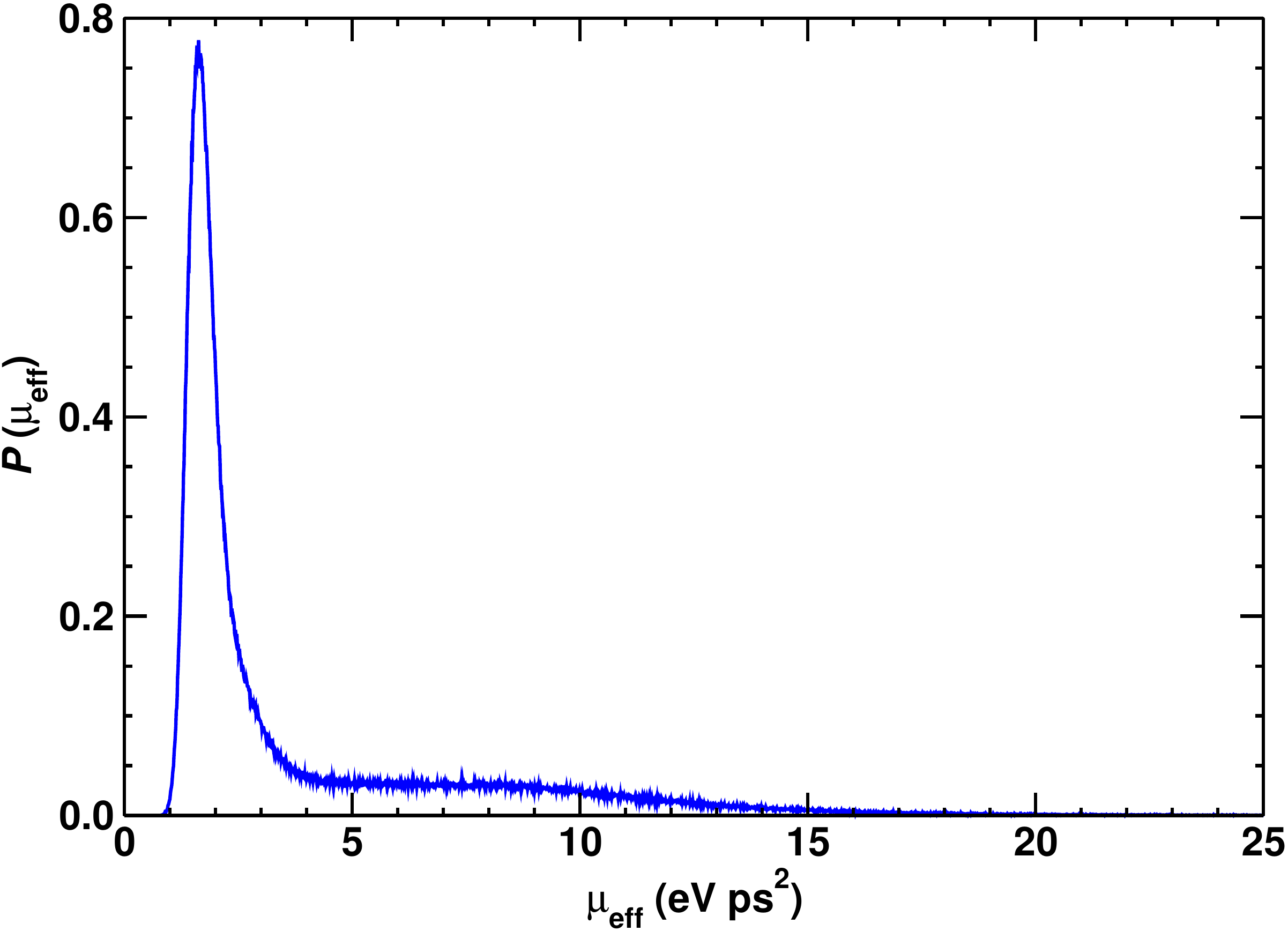} 
\end{center}
\caption{\label{fig:meff}
Distribution of the effective mass $\mu_{\text{eff}}$ extracted from 50 d-AFED runs.
}
\end{figure}

The d-AFED~\cite{Abrams2008} simulations were performed by implementing a corresponding \emph{fix} in the \textsc{lammps} code~\cite{lammps}.
To control the temperature of the physical system in the NVT simulations a Nos{\'e}-Hoover thermostat was applied as implemented in \textsc{lammps}.  A time step of $\Delta t=2$~fs was used in all simulations and the temperature of the physical system was set to $T=300$~K.  In each simulation the physical system was equilibrated for 1~ns before being coupled to the extended variable.  
For the extending variable the generalized Gaussian momentum thermostatting (GGMT)~\cite{Liu00} was used, with a temperature $T_s = 3000$~K, a mass $\mu_s = 4000$~eV$\cdot$ps$^2$, a coupling constant $\kappa = 2\times10^5$~eV, and the corresponding time constant $\tau = 2 \pi \sqrt{\mu_s/\kappa} = 0.89$~ps.
The effective mass of the 1D path CV, $f(\mathbf{Q}(\mathbf{r}))$, is given by~\cite{Cuendet14}
\begin{equation}
\label{eq:mueff}
\mu_{\text{eff}} = \left[ \sum_{j=1}^{N} \frac{1}{m_j} \left( \frac{d f}{d\mathbf{r}_j}   \right)    \right]^{-1} \quad ,
\end{equation}
where $N$ is the number of atoms, $\mathbf{r}_j$ are the atomic positions, and $m_j$ is their corresponding mass.  The effective mass of the path CV is not constant, but depends on the actual system configuration.  The distribution of effective masses extracted from 50 independent d-AFED runs is shown in Fig.~\ref{fig:meff}.  The distribution is clearly peaked around $\mu_{\text{eff}} = 1.6$~eV$\cdot$ps$^2$ and the average value is $\bar{\mu}_{\text{eff}} = 3.8$~eV$\cdot$ps$^2$, so that the ratio $\mu_s / \mu_{\text{eff}} \approx 1000$.
The typical time constant of the path CV is therefore $\tau_f = 2 \pi \sqrt{\bar{\mu}_{\text{eff}}/\kappa} \approx 0.028$~ps which is too short to be integrated with high accuracy using the MD time step $\Delta t = 2$~fs.
We therefore applied a dual time step algorithm that is comparable to the RESPA algorithms~\cite{Tuckerman92} with a time step of $\Delta t = 0.1$~fs.

\subsection{Metadynamics} 
The metadynamics simulations~\cite{Laio2002} were performed  by implementing a corresponding \emph{fix} in the \textsc{lammps} code.   The temperature and pressure was controlled by a Nos{\'e}-Hoover thermostat and barostat with Martyna-Tobias-Klein corrections~\cite{Martyna94} as implemented in \textsc{lammps}.  
In all simulations the physical system was initially equilibrated for 1~ns without any bias potential.
The time-dependent bias potential was build up by accumulating Gaussians
\begin{equation}
\label{metabias}
V^{\text{bias}} (s,t) = \sum_{t_i < t} h \exp\left( - \frac{(s - f(\mathbf{Q}(\mathbf{r}_{t_i})))^2}{2 \sigma^2}  \right) \quad .
\end{equation}
In the NVT simulations the height of the Gaussians was set to $h = 10^{-4}$~eV, the width to $\sigma = 0.005$, and an additional Gaussian was added with a frequency $\nu = 200$~fs.  The temperature was set to $T=300$~K.
NPT simulations were performed using $h = 0.002$~eV, $\sigma = 0.01$, and $\nu = 200$~fs with $T=300$~K and $P=0$.  In all simulations the bias potential was recorded on a grid with 1401 points for $s \in [-0.2,1.2]$.

\begin{figure}
\begin{center}
\includegraphics[width=0.44\textwidth]{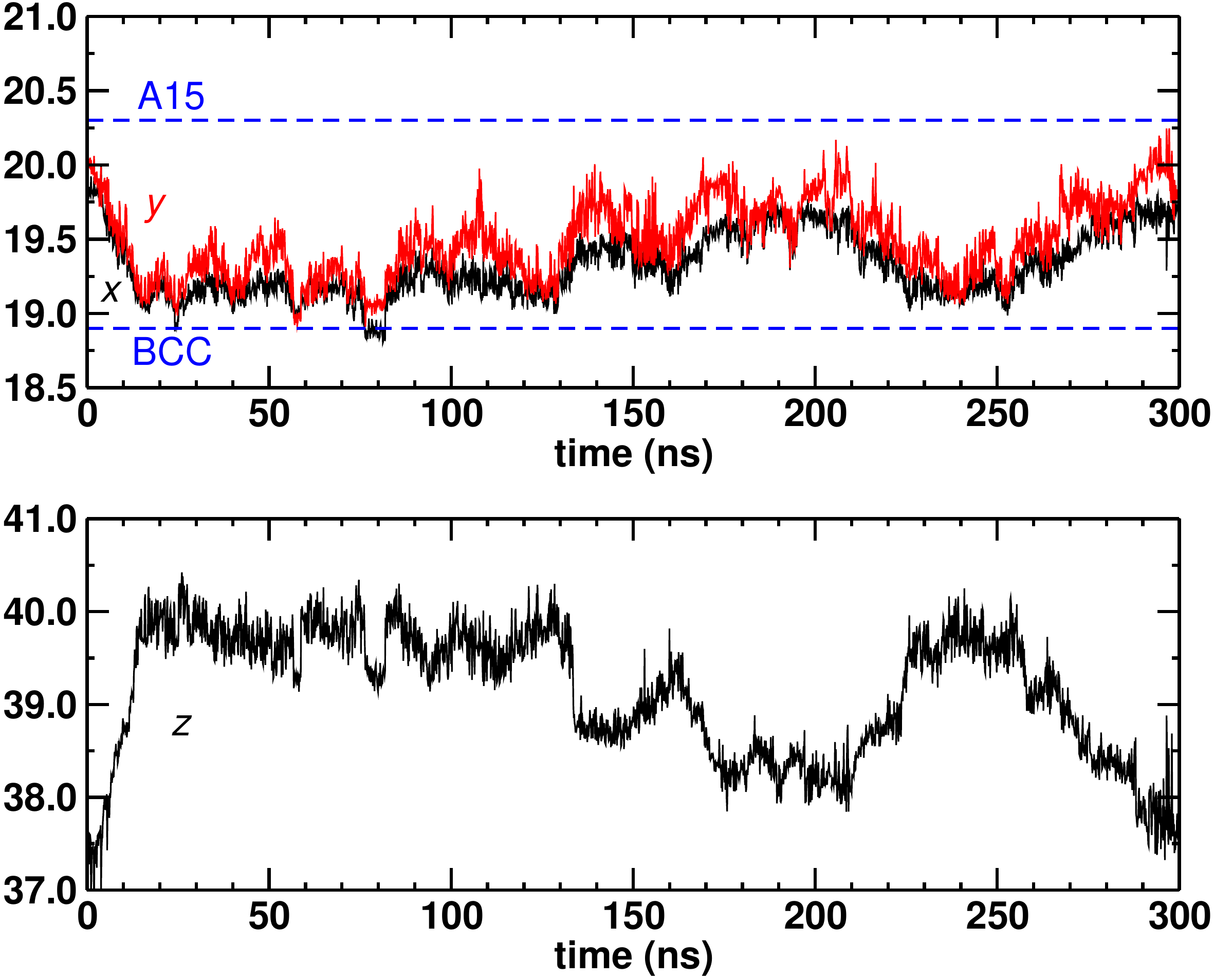} 
\end{center}
\caption{\label{fig:box}
Change in the lattice vectors during the NPT-metadynamics simulations.  As the phase fraction of bcc and A15 changes the dimensions adjust to the optimal values of the two phase.  Top: $x$- and $y$-dimension of the simulation cell. Bottom: $z$-dimension of the simulation cell.
}
\end{figure}

In the constant pressure simulations only the length of the lattice vectors was allowed to change while the cell shape was kept orthorhombic. The fluctuations in the lattice vectors are shown in Fig.~\ref{fig:box}.
As the phase fraction of bcc and A15 changes the dimensions of the simulation cell clearly adapts towards the optimal value of the corresponding phase.
